\documentclass[preprint,12pt]{elsarticle}

\usepackage{graphicx}
\usepackage{amssymb}
\usepackage{gensymb}
\usepackage{textgreek}
\usepackage[dvipsnames]{xcolor}
\usepackage{fontenc}
\usepackage[acronym]{glossaries}
\usepackage{siunitx}
\usepackage{url}
\usepackage{caption}
\usepackage{subcaption}
\usepackage[utf8]{inputenc}

\newacronym{aasi}{AaSI}{Aalto Spectral Imager}
\newacronym{radmon}{RADMON}{Radiation Monitor}
\newacronym{epb}{EPB}{Electrostatic Plasma Brake}
\newacronym{fmi}{FMI}{Finnish  Meteorological  Institute}
\newacronym{eps}{EPS}{Electrical Power Subsystem}
\newacronym{tt&c}{TT\&C}{Telemetry, Telecommand \& Communication}
\newacronym{adcs}{ADCS}{Attitude Determination \& Control Subsystem}
\newacronym{leop}{LEOP}{Launch \& Early Operations Phase}
\newacronym{eet}{EET}{Eastern European Time}
\newacronym{obdh}{OBDH}{Onboard Data handling}
\newacronym{obc}{OBC}{On Board Computer}
\newacronym{gui}{GUI}{Graphical User Interface}
\newacronym{mcc}{MCC}{Mission Control Center}
\newacronym{eqm}{EQM}{Engineering-Qualification Model}
\newacronym{fm}{FM}{Flight Model}
\newacronym{vtt}{VTT}{VTT Technical Research Centre of Finland}
\newacronym{fpi}{FPI}{Fabry--P\'{e}rot Interferometer}
\newacronym{pcb}{PCB}{Printed Circuit Board}
\newacronym{uhf}{UHF}{Ultra High Frequency}
\newacronym{gps}{GPS}{Global Positioning System}
\newacronym{epscb}{EPSCB}{Electrical Power System Control Board}
\newacronym{fmea}{FMEA}{Failure mode and effects analysis}
\newacronym{sdr}{SDR}{Software Defined Radio}
\newacronym{tle}{TLE}{Two Line Element}
\newacronym{sdram}{SDRAM}{Synchronous Dynamic Random Access Memory}
\newacronym{i2c}{I$^2$C}{Inter Integrated Communication}
\newacronym{uart}{UART}{Universal Asynchronous Receiver Transmitter}
\newacronym{spi}{SPI}{Serial Peripheral Interface}
\newacronym{pate}{PATE}{Particle Telescope}
\newacronym{leo}{LEO}{Low Earth Orbit}
\newacronym{cots}{COTS}{Commercial Off The Shelf}
\journal{Acta Astronautica}

\begin{document}

\begin{frontmatter}

\title{Aalto-1, multi-payload CubeSat: In-orbit results and lessons learned}

\author[1]{M.~Rizwan~Mughal\footnote{M. Rizwan Mughal is also associated with Electrical Engineering Department, Institute of Space Technology, Islamabad, Pakistan, Correspondence: rizwan920@gmail.com}
}
\author[1]{J.~Praks}
\author[2]{R.~Vainio}
\author[3]{P.~Janhunen}
\author[3]{J.~Envall}
\author[4]{A.~N\"asil\"a}
\author[2]{P.~Oleynik }
\author[1]{P.~Niemel\"a}
\author[1,5]{A.~Slavinskis}
\author[2]{J.~Gieseler}
\author[1]{N.~Jovanovic}
\author[1]{B.~Riwanto}
\author[3]{P.~Toivanen}
\author[1]{H.~Leppinen}
\author[1]{T.~Tikka}
\author[2]{A.~Punkkinen}
\author[6]{R.~Punkkinen}
\author[6]{H.-P.~Hedman}
\author[7]{J.-O. Lill}
\author[8]{J.M.K. Slotte}

\address[1]{Department of Electronics and Nanoengineering, Aalto University School of Electrical Engineering, 02150 Espoo, Finland}
\address[2]{Department of Physics and Astronomy, University of Turku, 20014 Turku, Finland}
\address[3]{Finnish Meteorological Institute, Space and Earth Observation Centre, Helsinki, Finland}
\address[4]{VTT Technical Research Centre of Finland Ltd, Espoo, Finland}
\address[5]{Tartu Observatory, University of Tartu, Observatooriumi 1, 61602 T{\~o}ravere, Estonia}
\address[6]{Department of Future Technologies, University of Turku, 20014 Turku, Finland}
\address[7]{Accelerator Laboratory, Turku PET Centre, \AA{}bo Akademi University, 20500 Turku, Finland}
\address[8]{Physics, Faculty of Science and Technology, \AA{bo} Akademi University, 20500 Turku, Finland}

\begin{abstract}
 
The in-orbit results and lessons learned of the first Finnish satellite Aalto-1 are briefly presented in this paper. Aalto-1, a three-unit CubeSat which was launched in June 2017, performed \gls{aasi}, \gls{radmon} and \gls{epb} missions. The satellite partly fulfilled its mission objectives and allowed to either perform or attempt the experiments. Although attitude control was partially functional, \gls{aasi} and \gls{radmon} were able to acquire valuable measurements. \gls{epb} was successfully commissioned but the tether deployment was not successful.

In this paper, we present the intended mission, in-orbit experience in operating and troubleshooting the satellite, an overview of experiment results, as well as lessons learned that will be used in future missions.

\end{abstract}


\end{frontmatter}

\section{Introduction}
\label{S:1}

There has been a significant increase in the design, development, launch and operation of nano and micro satellites since last two decades. A large number countries initiated their space activities and a large number of Newspace companies emerged as an outcome. A number of innovative platform subsystems, payloads and missions have been proposed, designed and launched by universities and small industry thanks to significantly reduction of development and launch costs \cite{9183926,7015716,Ali2018,Ali2013INNOVATIVEEP,mukhtar,mughal,mughal_intra,ali2014innovative,mughal2014plug}. This has been made possible due to availability of \gls{cots}, technology miniaturization and affordable rides.
The CubeSat standard, initially perceived for educational purposes only, was defined by Stanford and California Polytechnic State Universities in 1999 \cite{Bouwmeester2010}. Since the launch of first CubeSat in 2003, this standard has revolutionized the space industry by playing an increasingly important role in technology demonstrations, remote sensing, Earth observation and education \cite{Crusan2019,Poghosyan2017}. More recently, the CubeSats have started to increasingly exploit the scientific and commercial use cases \cite{Poghosyan2017,SEYEDABADI}. Being small in size, they have transformed the traditional design approach of space systems by providing low-cost access to space \cite{Frischauf2018,Peters2015,Tkatchova2018,Salt2013}. A single ride of launch vehicle can carry hundreds of CubeSat-class satellites. Many universities are effectively using CubeSats as hands-on tools to teach the challenging engineering concepts about the design and development of complex interdisciplinary systems. The launch and operation phase provides a unique learning experience to university teams enabling them to learn essential skills in mission design and operations \cite{Zurbuchen2016}. Now a day's university CubeSat missions aim at real science and technology demonstration while also ensuring the educational objectives. It is important for CubeSat community to share the knowledge, in orbit experiences, lessons learned and mission details which will consequently help other teams to gain valuable experience and not repeat the same mistakes.

The current small satellite literature lacks the whole life cycle: i.e. all aspects relating to mission planning, design, launch, operations and lessons learned. The teams either report very specific technical information of the design, or come up with mission descriptions and in-orbit results. One can barely find information in the current literature about complete life cycle covering a wide range of aspects. In order to provide the CubeSat community with the sufficient details on complete aspects in terms of technology development, technology demonstration and key experiences, we present the design, development and in-orbit experience of Aalto-1, the first satellite of Aalto University, Finland. We present our findings in two papers: the first one covering the technology development aspects \cite{a1_technology_2020} whereas the present paper covers the in-orbit results and lessons learned.

\section{Mission overview}
\label{overview}
\begin{figure}[h!]
    \centering
    \includegraphics[width=\columnwidth]{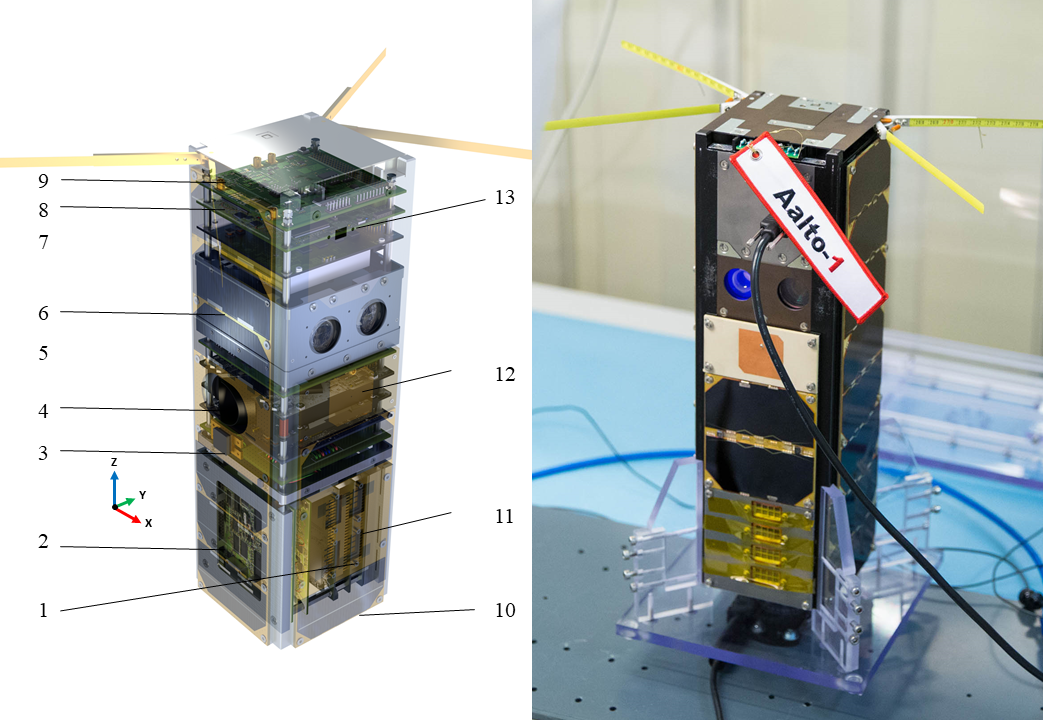}
    \caption{Overview of Aalto 1 subsystems and photograph of \gls{fm}.The highlighted subsystems are: 1) Radiation Monitor (RADMON), 2) Electrostatic Plasma Brake (EPB) 3) Global Positioning System's (GPS's) antenna and stack interface board, 4) Attitude Determination and Control System (ADCS), 5) GPS and S-band radio, 6) Aalto Spectral Imager (AaSI), 7) Electrical Power System (EPS), 8) On-Board Computer (OBC), 9) Ultra High Frequency (UHF) radios, 10) solar panels, 11) electron guns for EPB, 12) S-band antenna, 13) debug connector}
    \label{fig:a1_photo}
\end{figure}

Aalto-1, shown in Fig. \ref{fig:a1_photo}, is a 3U CubeSat designed and developed by Aalto university and partner organizations. The spacecraft was launched in June 2017 and hosted three payloads: \gls{aasi}, \gls{radmon} and \gls{epb}. 

\gls{aasi} is the first hyperspectral imaging system compatible with nanosatellites, based on a piezo-actuated tunable \gls{fpi} which allows for an unprecedented miniaturization~\cite{Praks2018MiniatureSI}. The instrument fits in a half of CubeSat unit and, within a few seconds, can acquire spectral images in tens of freely programmable channels. The filter works in the spectral range of 500--900~nm where each channel is 10--20~nm wide. A 512$\times$512-pixel sensor with a 10$^{\circ}$ field of view provides a ground resolution better than 200~m per pixel.

\gls{radmon}, fitting within 0.4 CubeSat units, is one of the smallest particle detectors, which has proven itself capable of taking scientific measurements~\cite{Gieseler2020Radmon, Oleynik-etal-2020}. It measures electron energies in the $>$1.5~MeV range and protons in the $>$10~MeV range.

\gls{epb} is novel deorbiting technology which employs the coulomb drag between the ionospheric plasma and a long charged tether~\cite{epb_main, eas_A1}. The tether is deployed using a centrifugal force and it is estimated that a 100-m tether (such as on-board Aalto-1) could decrease an altitude by 100 km of a three-unit CubeSat within 600 days~\cite{Iakubivskyi2019}. A similar experiment was carried on-board ESTCube-1~\cite{eas_ec1, eas_pl} where tether deployment was not successful~\cite{estcube1_lessons}. While Aalto-1 \gls{epb} experiment was improved based on ESTCube-1 ground test results, yet the deployment of \gls{epb} was not successful. This is due to the fact that Aalto-1 flight hardware had to be delivered soon after the ESTCube-1 experiment was carried out and, therefore, the team did not have time and resources to redesign the \gls{epb} module, as it is being done for the FORESAIL-1 mission~\cite{Iakubivskyi2019}. 

In this paper, section~\ref{sec:timeline} briefly introduces mission timeline representing the launch and operations. Section~\ref{sec:paylaods} presents the in-orbit results and lessons learned of all the payloads. \gls{radmon} in-orbit results are introduced briefly based on previously published results~\cite{Gieseler2020Radmon, Oleynik-etal-2020}. \gls{epb} in-orbit results and lessons learned are discussed in detail, especially the possible reasons of tether deployment failure. \gls{aasi} detailed in-orbit results are presented here for the first time. Furthermore, Section~\ref{sec:platform} introduces in-orbit experience of platform's subsystems. Section~\ref{sec:discussion} discusses the results and concludes the paper.

\section{Mission timeline}
\label{sec:timeline}

The spacecraft was launched aboard PSLV-C38 launch vehicle at 05:59 \gls{eet} and the first beacon was recorded by a \gls{sdr} located in South Africa at approximately 08:30 \gls{eet}. The first contact with the Aalto University ground station was established during the first pass at 10:07 \gls{eet}. 

During the consequent passes, several responses were recorded, but were not decoded due to an unidentified problem in the ground station reception chain. Later on the problem was troubleshooted  to be in mast pre-amplifier.  While powering it off provided a directional link with the CubeSat, it came at a cost -- a loss in the signal strength.

\begin{figure}[h!]
    \centering
    \includegraphics[width=\columnwidth]{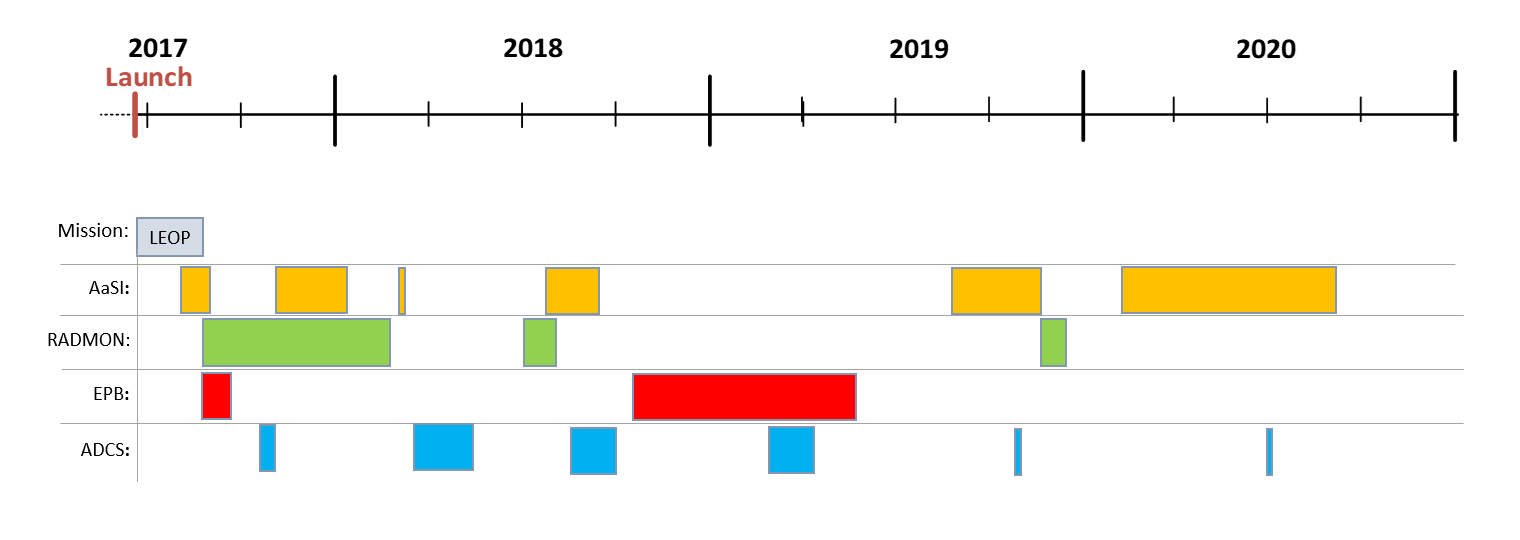}
    \caption{Mission timeline}
    \label{fig:a1_timeline}
\end{figure}

The mission wise timeline on the commissioning and  operations of each experiment is presented in Fig.~\ref{fig:a1_timeline}. During \gls{leop}, the first \gls{aasi} picture was downloaded and \gls{radmon} commissioning phase was started. As part of Aalto-1 operations, multiple \gls{aasi} campaigns have been completed.
\gls{radmon} operations resulted in a useful data set during nominal conditions and also during a solar storm. \gls{epb} campaign resulted in partial success in commissioning phase but failure in tether deployment.

\section{Mission payloads}
\label{sec:paylaods}
This section describes the in orbit performance of \gls{radmon}, \gls{epb} and \gls{aasi} payloads. The thorough design approach, selection and implementation has been presented in accompanying paper \cite{a1_technology_2020}.

\subsection{Radiation monitor mission}
\begin{figure}[h!]
\centering
\includegraphics[width=3cm]{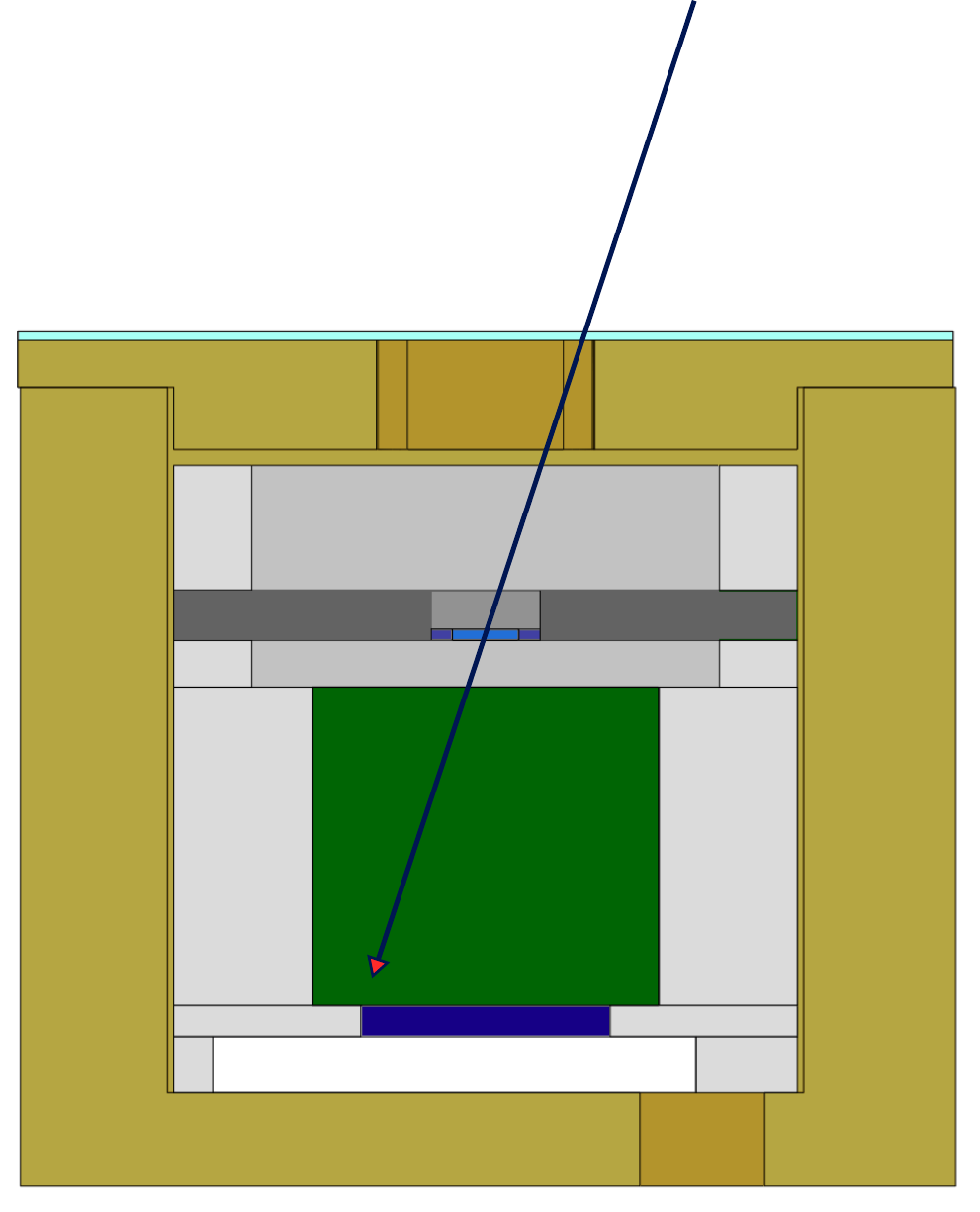}
\caption{The \gls{radmon} radiation monitor cross section. The arrow on the picture shows a particle that is incident within the instrument aperture. The brass case is light-brown. The silicon detector is light blue, surrounded by a blue passive silicon area, which is fixed on a printed circuit board (PCB) shown as dark gray. The CsI(Tl) scintillator is shown in green. Under the scintillator there is a photodiode shown in dark blue. White structure on the bottom is an alumina case of the photodiode.}
\label{fig:radmoncross}
\end{figure}

The \gls{radmon} is a small (4$\times$9$\times$10~$\text{cm}^3$, 360 g) low-power (1 W) radiation monitor \citep{Peltonen2014, a1_technology_2020}. The monitor detects protons and electrons employing a regular \textDelta{E} -- E analysis to distinguish between particle species. The detectors of the instrument are a 2.1$\times$2.1$\times$0.35~$\text{mm}^3$ silicon detector and a 10$\times$10$\times$10~$\text{mm}^3$ CsI(Tl) scintillation detector placed inside a brass envelope (see Figure \ref{fig:radmoncross}). The envelope of the detector compartment is opaque for protons below 50 MeV and electrons below 8 MeV.
The envelope has a 280 $\mathrm{\mu m}$ aluminum entrance window that stops low energy photons and low energy charged particles. A particle must hit both detectors to be registered. Therefore, the thicknesses of the entrance window and the silicon detector set the lower energy threshold for protons to about 10 MeV and electrons to about 1.5 MeV. A detailed description of the instrument calibration is presented in \cite{Oleynik-etal-2020}.

\subsubsection{In-orbit results}
\gls{radmon} in-orbit calibration campaign was carried out in September 2017. It was discovered that the gain of the scintillator did not match the value obtained from ground calibrations, but was about 30\% lower. The reason could not be positively determined, but the deterioration of the optical contact between the CsI(Tl) crystal and the photodiode during launch vibrations could potentially be responsible for this decay of performance. A successful in-flight calibration was, however, achieved using data obtained in a dedicated calibration mode, which allows raw data from detectors to be down-linked. The in-flight calibration is discussed in detail in \cite{Oleynik-etal-2020}.

\begin{figure}[h!]
\centering
\includegraphics[width=0.7\textwidth]{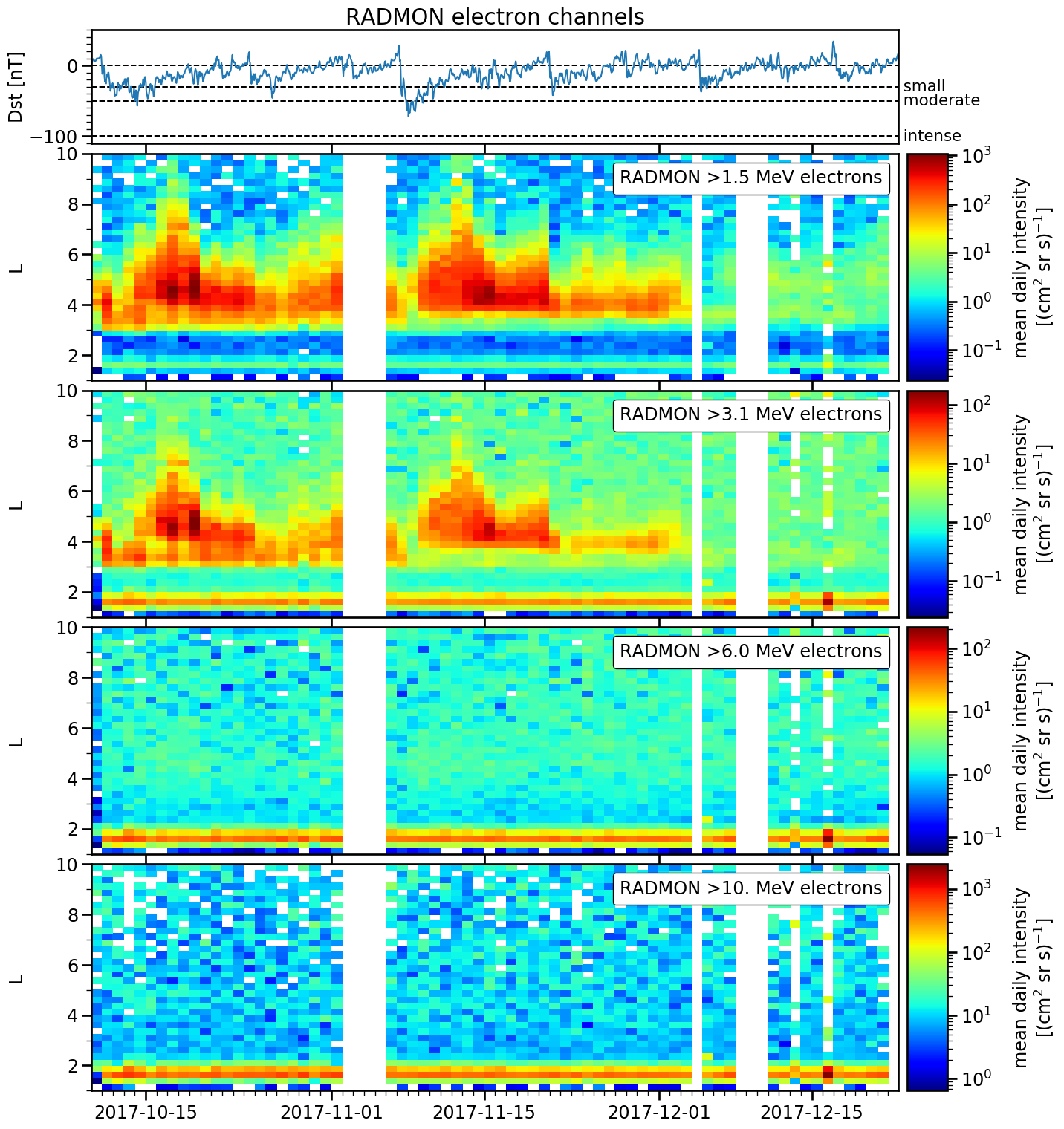}
\caption{From top to bottom: Time series of $Dst$ index and four histograms of integral intensities with respect to $L$ parameter obtained by the different \gls{radmon} electron channels from 10 Oct 2017 to 21 Dec 2017. The z-axis gives color-coded arithmetic daily mean of intensity per bin -- note that the color scale is different for all panels in order to enhance the details of all channels which have different sensitivities. Figure adapted from \cite{Gieseler2020Radmon} by permission of Elsevier, \copyright2019 COSPAR.}
\label{fig:ldynamic}
\end{figure}

The first observational campaign of \gls{radmon} started on 10 October 2017 and lasted until 2 May 2018. Using these data, it has been demonstrated in \cite{Gieseler2020Radmon} that the instrument is able to measure the integral intensities of electrons above 1.5~MeV and protons above 10~MeV in \gls{leo}, reflecting the dynamic environment of the radiation belts. Fig.~\ref{fig:ldynamic} shows the temporal evolution of daily electron intensities from October to December 2017 with respect to McIlwain $L$ parameter \cite{McIlwain1961} (indicating the equatorial distance of drift shells) together with the $Dst$ (disturbance storm time) index as a measure for geomagnetic storm intensity \cite{Gonzalez1994,Pulkkinen2007}. The two observed moderate geomagnetic storms result in strong enhancements of the outer radiation belt, while periods following small storms are characterized by reduced electron intensities in the outer belt.

Figure~\ref{fig:ldynamic} also illustrates the contamination of all electron measurements by higher energy protons: the constantly increased intensities in the $L$ range below 2 correspond to the proton-dominated inner radiation belt.
Further comparisons with electron spectra observed in a similar but slightly higher orbit (820~km) by the Energetic Particle Telescope (EPT) onboard the ESA minisatellite (volume $<$1~m$^3$) PROBA-V (PRoject for OnBoard Autonomy-Vegetation) showed a good agreement for the $>$1.5~MeV electron channel of \gls{radmon} \cite{Gieseler2020Radmon}.

Next observation effort was made late in 2019 to check if the instrument functions well.  We have ensured that the instrument is in a good shape, but the satellite lacks power for continuous operations of \gls{radmon}. A compromise was found to keep \gls{radmon} operating for every 12 hours with a 3-hour break to ensure recharge of the satellite battery. A new set of calibration data from the end of 2019 confirmed that the calibration of the detectors had not changed during the 2.5 years in space and that no visible signs of detector degradation could be identified.

\subsubsection{Lessons learned}
\gls{radmon} is a successful space experiment and, certainly, it can be improved. Minimization of the contamination of electron channels by high-energy protons would be the most valuable improvement for the instrument. The collimator geometry should also be streamlined to achieve an optimal instrument aperture.

The current design is such that particles enter the instrument within a $\approx 20 \degree$ half-width cone defined by an opening in the brass container. The opening is manufactured as a right-angle shaft sufficiently larger than the dimension of the silicon detector (see Fig.~\ref{fig:radmoncross}).

An incident particle may, therefore, hit a side of the silicon detector in a way that it deposits energy into its active area and its passive area in an arbitrary proportion. Further, it hits the scintillation detector. This effect leads to an underestimation of energy deposited in the \textDelta{E} detector. Subsequently, such a particle is misclassified. A silicon detector with two concentric active areas would contribute to better particle classification and reduce contamination of electron channels by protons. The detector should trigger on the central dot and add the energy deposited in the encircling area to its output signal. One of the possible geometries could be a "sandwich" detector with a thinner layer carrying the central spot and a thicker layer beneath. In this case, it is even easier to get the correct \textDelta{E} signal since the energy loss in the thinner layer would not be needed for the pulse height analysis. Any signal above the threshold would gate the particle detection by the \textDelta{E} -- E detectors below. The thickness of the top detector can be about 100--150 $\mathrm{\mu}$m and should be optimized for scientific requirements. A thicker detector would show more edge effects than a thin one, but could have a better signal-to-noise ratio. The thickness of the entrance window should be adjusted as well during the optimization.

Another issue is that the current geometry allows a gradual increase in the angle of the acceptance cone. The collimator should be designed as a conical opening in the shielding container so that it becomes transparent at sharper energy threshold. It would improve the flatness of the particle response at moderate energies.

\begin{figure}[h!]
\centering
\includegraphics[width=0.8\textwidth]{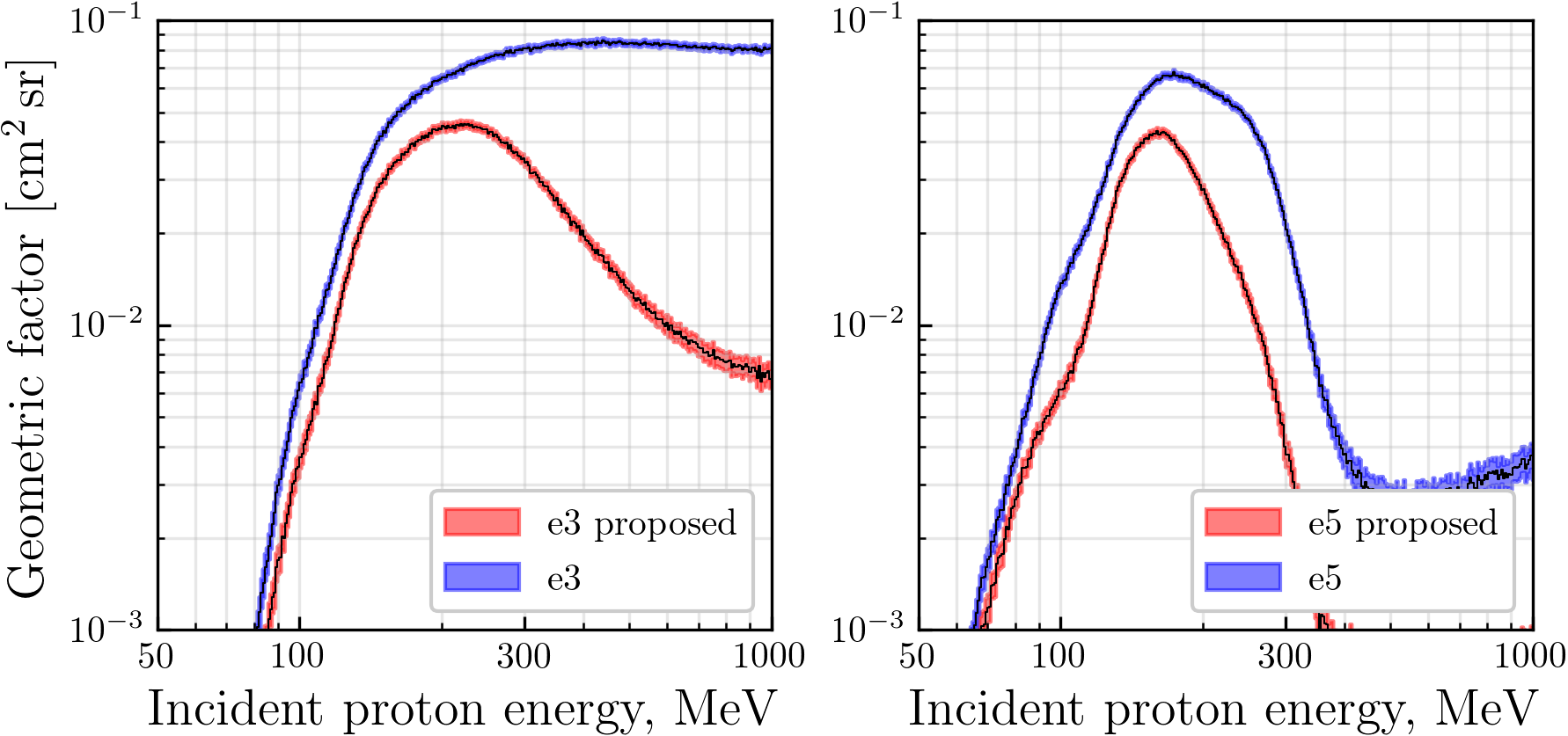} 
\caption{The contamination of electron channels (e3 and e5 are chosen as examples) by high energy protons in comparison to a proposed "sandwich" design.}
\label{fig:radmonextra}
\end{figure}

A simulation of the suggested layered design carried out within the Geant4 \citep{GEANT4-AGOSTINELLI, GEANT4-Allison} framework  is compared to a simulation of the current design in Fig.~ \ref{fig:radmonextra}. The "sandwich" has a thin silicon detector right on top of the \textDelta{E} silicon detector. Both detectors are square and of the same size. The instrument container has a tantalum front wall, which can be optimized further to a tantalum lining of the container opening. This reduction is possible since the upper thin silicon detector sets the accepted solid angle for a particle to be detected. High energy protons coming within the aperture are still detected as electrons. Nevertheless, limiting the solid angle of the instrument acceptance for such protons improves the quality of the observational data. In a proton-rich environment, such as South Atlantic geomagnetic anomaly, contamination of electron channels could be used as a secondary proxy on the proton population.

As a positive takeaway from the experiment, the successful \gls{radmon} re-calibration using in-flight data showed that a dedicated mode allowing the full pulse height data to be downloaded also from space can render a \gls{radmon}-like instrument to a self-calibrating device. Thus, an expensive full calibration campaign in high-energy beam facilities, reaching hundreds of MeVs in proton energies, can be avoided using this approach.

\subsection{Electrostatic Plasma Brake}
\begin{figure}[h!]
    \centering
    \includegraphics[width=\columnwidth]{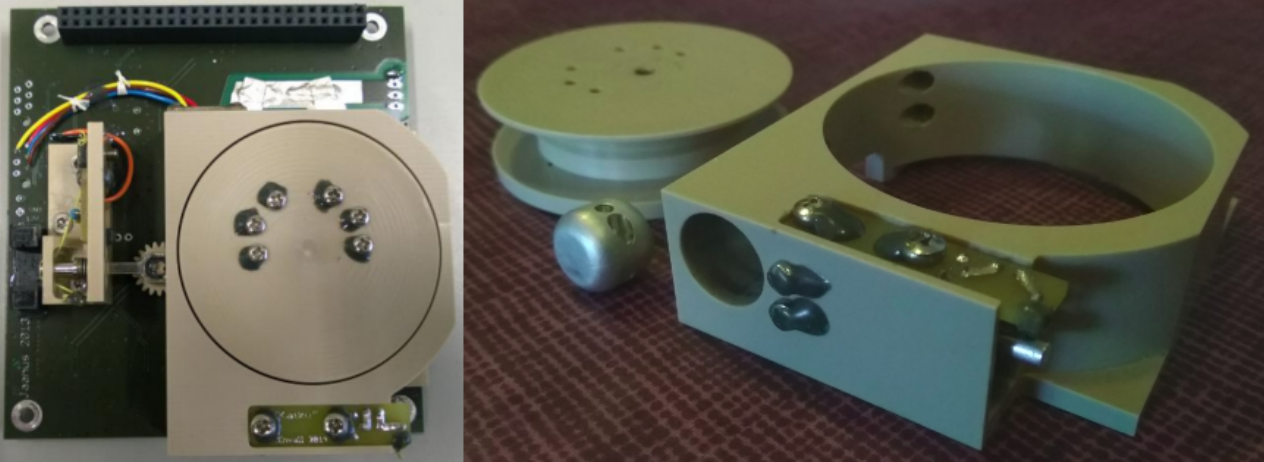}
    \caption{\gls{epb} Mechanical parts on the \gls{pcb} (left) and key tether reeling components: reel, tip mass, tether chamber, and tip mass launch lock (right). The reel lock can be seen on the left panel left side of the tether chamber.}
    \label{fig:epbmechanics} 
\end{figure}

The key components of the \gls{epb} payload are those of the tether reeling mechanism as shown in Fig~\ref{fig:epbmechanics}. These include the tether reel, reel motor (not visible), tether chamber, tether tip mass, tip mass launch lock (Kaiku), and reel launch lock (Kieku). The reel motor (vacuum qualified piezo motor) is nested inside the reel. The control electronics underneath the tether reeling \gls{pcb}, separate high voltage \gls{pcb}, and electron emitters are similar to those of ESTCube-1 as described earlier in the literature~\cite{eas_pl}. Only changes introduced were related to the revised launch locks and additional diagnostics. The high voltage converter was changed to double the voltage from $\pm$ 500 kV to $\pm$ 1000 kV which caused some minor changes in the electron emitters.

The revised launch locks and additional diagnostics included the following components. The reel lock was newly designed and diagnostics was added. Behind the spring loaded lock shaft is an optoport (black component next to the lock in the left panel of Fig.~\ref{fig:epbmechanics}. When the lock was burned the state of the optoport was designed to change from open to close. To monitor the tip mass before and after the tip mass lauch lock was released, a pair of phototransistor (Kyyl\"a) and IR LED (Soihtu) was mounted in the tether chamber opposite to the opening tube of the tip mass and tether (two holes in the top right corner of the tether chamber in the right panel of Fig.~\ref{fig:epbmechanics}). The IR LED can be used to healty check the phototransistor prior to the tip mass release. After the release, if the tether should be damaged, the phototransistor observed light freely entering to the tether chamber.

\subsubsection{In-orbit results}

The in-orbit tests of the plasma brake started with a commission phase, in which the \gls{obc} sent \gls{epb} a number of commands with the goal of verifying its operational state. This list included essentially all the commands which were safe to run without any risk of hazard. This restriction ruled out e.g. the commands that would initiate physical changes in the payload's status (launch lock burns, motor activation) or the ones not usable at this point of the mission (high voltage or electron gun activation). The commands that were run all worked as designed, returning some housekeeping data for analysis. Most importantly at this point, the data showed that all systems were at nominal state and that the launch locks had kept the tether reel and the tip mass intact.

\begin{figure}[h!]
    \centering
    \includegraphics[width=\columnwidth]{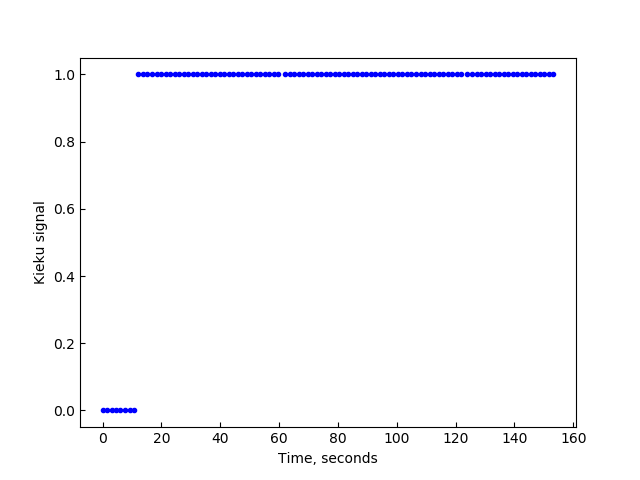}
    \caption{Flight data showing the deployment of the tether reel lock Kieku.}
    \label{fig:kieku}
\end{figure}

The second step in in-orbit tests was to open the two launch locks that had locked the tether reel and the tip mass during the launch. Each lock was opened by applying a 150 mA current, which would melt the dyneema string keeping the spring loaded lock at closed state. The tether reel lock, named Kieku, had an integrated optical diagnostics system whose state could be read by the OBC at any time. During the burn sequence the system's state switched from ``locked'' to ``deployed'' after about 12 seconds of burning as shown in Fig.~\ref{fig:kieku}. Similar diagnostics were not available for the tip mass lock Kaiku. The duration of the burn current for Kaiku was chosen long enough to ensure a proper deployment.

\begin{figure}[h!]
    \centering
    \includegraphics[width=\columnwidth]{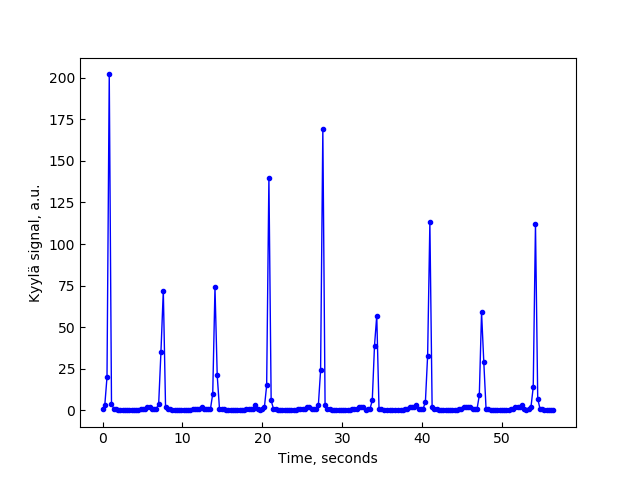}
    \caption{Flight data from the phototransistor Kyyl\"a. The periodicity of the signal corresponds to the satellite's spin rate at the time.}
    \label{fig:kyyla}
\end{figure}

The final verification of the operational readiness before attempting tether deployment was performed with the help of the photosensor Kyyl\"a. Kyyl\"a is a simple phototransistor placed inside the tether reel chamber. It has the backside of the tip mass in the center of its field of view. If the tip mass had been ejected from its nest prematurely, the light (e.g. from Earth albedo) entering the chamber could easily be detected in Kyyl\"a's signal. An example data plot from an early Kyyl\"a scan is shown in Fig.~\ref{fig:kyyla}. The extremely narrow width of the peaks indicate that even though light is able to enter the chamber, it is able to do so over a very narrow angle only, as the satellite is spinning. This may be explained as follows. The tip mass, roughly cylinderic, remains like a plug in the tether opening tube. The tip mass is not tightly in the tube but held to its place by the launch lock. Thus there is a tiny gap between the tip mass and the tube walls that provides the light with a passage of the narrow angle. If the tip mass was completely removed, the shape of the peaks would be considerably wider. Another piece of information obtained from these tests was the confirmation of the satellite's spin rate. An approximate seven second periodicity of the peaks coincided precisely to the angular velocity data of the \gls{adcs}. Simultaneously it provided proof that Kyyl\"a was indeed measuring real phenomena of its surroundings and not some arbitrary electrical disturbances.

After the successful initial tests and preparations it was time to attempt tether deployment. A controlled spin-up of the satellite could not be performed due to the shortcomings of the satellite's \gls{adcs} which are described in detail in section\ref{sec:platform}. The satellite was nonetheless spinning through natural causes and its spin axis and angular velocity ($\approx$50 degrees per second) were, by chance, suitable for taking a shot at deployment. 

\begin{figure}[h!]
    \centering
    \includegraphics[width=\columnwidth]{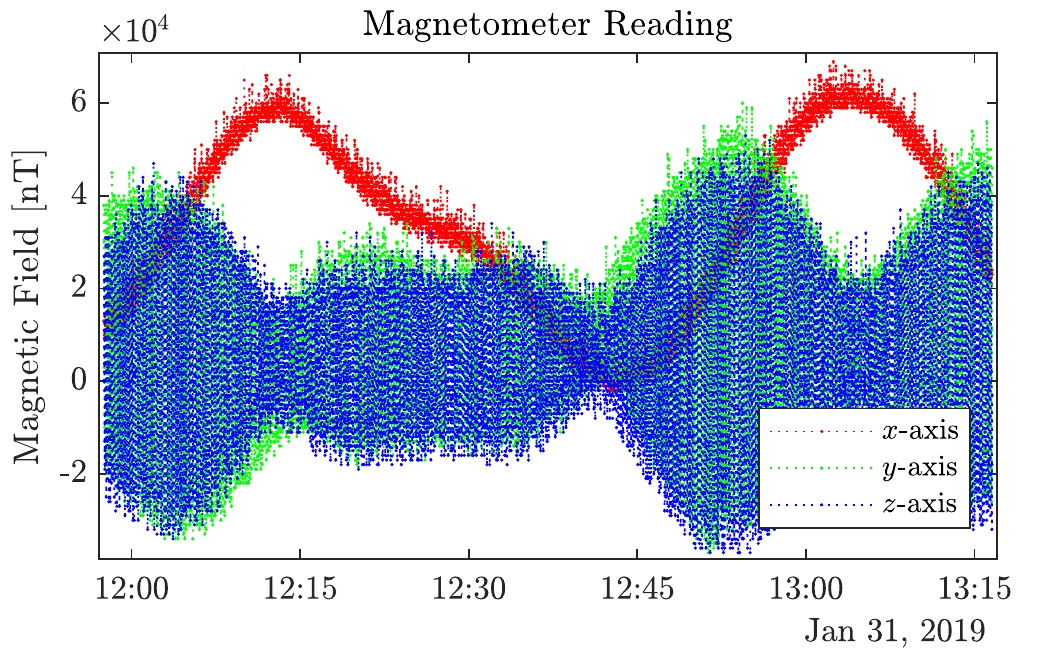}
    \caption{Magnetometer high resolution data during \gls{epb} deployment campaign}
    \label{fig:mgm_time}
    \end{figure}

\begin{figure}[h!]
    \centering
    \includegraphics[width=\columnwidth]{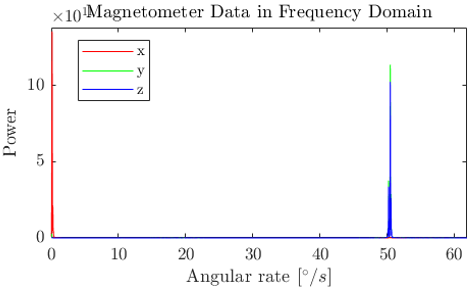}
    \caption{Magnetometer high resolution flight data during \gls{epb} deployment campaign in frequency domain confirming the spin rate around the spin axis}
    \label{fig:mgm_freq}
\end{figure}

\begin{figure}[h!]
    \centering
    \includegraphics[width=\columnwidth]{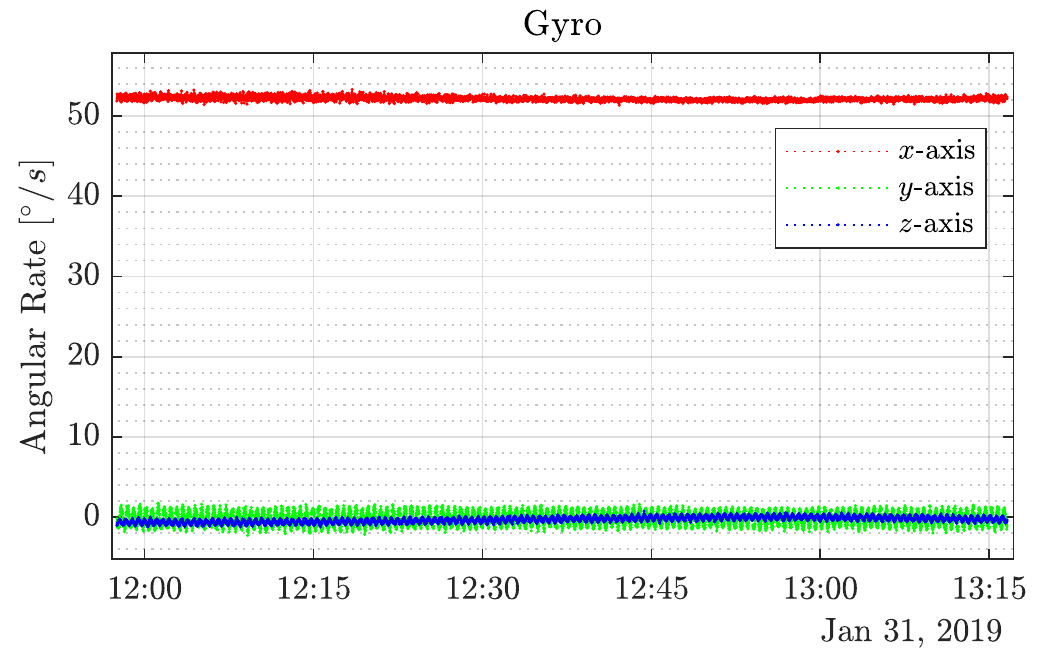}
    \caption{High resolution gyroscope calibrated flight data during \gls{epb} deployment campaign}
    \label{fig:gyro_data}
\end{figure}

The spin rate for \gls{epb} deployment was verified by magnetometer and gyroscope telemetry data. Figures~\ref{fig:mgm_time} and \ref{fig:mgm_freq} present the high resolution measurement data in time and frequency domains respectively and Fig.~\ref{fig:gyro_data} presents the calibrated gyroscope data during the \gls{epb} deployment campaign. 

\begin{figure}[h!]
    \centering
    \includegraphics[width=\columnwidth]{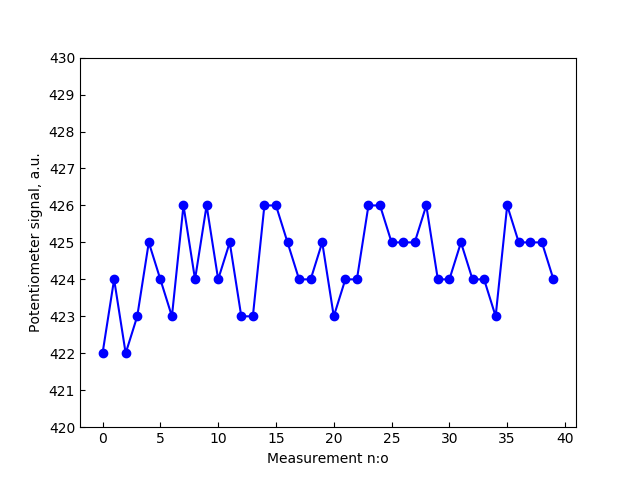}
    \caption{Measured values of the rotary encoder of the tether reel motor. These values were recorded over several tether deployment attempts. The peak-to-peak variation of the values corresponds to a turn of 1.4 degrees. The pre-launch value recorded in the last ground tests was 421.}
    \label{fig:rotary_encoder}
\end{figure}

Despite having achieved the desired spin rate around tether deployment axis, the deployment attempts all failed, unfortunately. In each attempt the motor was commanded to make a turn that is relatively small but still easily detectable. We couldn't observe any changes in the tether reel rotary position. The vacuum qualified piezo motor has an in-built potentiometer based rotary encoder.  Fig.~\ref{fig:rotary_encoder} shows the values measured by this encoder throughout the tether deployment trials. The peak-to-peak variation of the values corresponds to a 1.4{\degree} turn, or 0.4~mm on the perimeter of the reel. The conclusion must be that no detectable motor movement has taken place. If the motor had worked nominally, the turn angle would have been tens of degrees.

\subsubsection{Lessons learned}
The most noticeable result of the \gls{epb} mission is obviously the failure of the tether deployment hardware. It is somewhat unclear why this happened, even though several clues exist. Figure~ \ref{fig:rotary_encoder} shows examples of the measured motor voltage during two tether deployment attempts. In normal operation the motor voltage would remain in its nominal value of approximately 40~volts. As the plots show, the voltage is cut off and starts a rather rapid decay as soon as it has been switched on. The motor voltage is generated within the \gls{epb} control electronics with the help of a boost converter. In Fig.~\ref{fig:rotary_encoder} the voltage appears to saturate at the level of the boost converter's input voltage. This would indicate that the faulty operation of the boost converter is the source of all grief.

\begin{figure}[h!]
    \centering
    \includegraphics[width=\columnwidth]{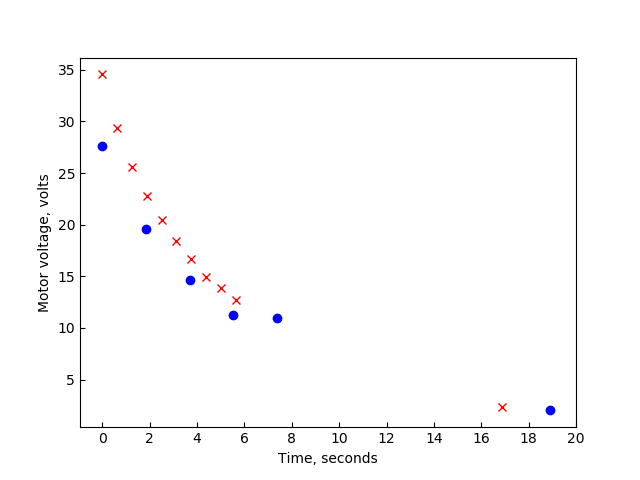}
    \caption{Two data sets of the motor voltage during the tether deployment attempts. Notice the saturation at the level of the input voltage ($\approx$11 volts) of the boost converter. In each set the last data point was recorded after the input voltage had been switched off.}
    \label{fig:motor_voltage}
\end{figure}

Not all went haywire, though. Several newly developed systems, some including moving parts, worked exactly as planned. Especially the completely renewed design for the tether reel lock Kieku proved to be a reliable work horse in space. At this point it is important to introduce the reader to the launch history of the \gls{epb} payload. A very similar payload was first launched on-board ESTCube-1 \cite{eas_ec1, eas_pl}. Its fate was identical to that of Aalto-1 \gls{epb}. It is important to note that the timelines of the two satellite missions overlapped in a most unfortunate way. Once the in-orbit results of ESTCube-1 were ready and verified, the delivery date of the Aalto-1 flight model hardware was only four months away. Also, due to the lack of proper on-board diagnostics, the reasons for the failure were mostly unknown. Therefore the \gls{epb} team was lacking both the proper time and the accurate knowledge of the problem in order to make fundamental changes in the motor hardware and control electronics. Instead, a number of features were added to gather all the information possible, in order to at least see what is happening in case of repeated failure. All these diagnostics tools described above (Kyyl\"a, Kieku's optical feedback, motor's position encoder) worked as planned. This allowed the \gls{epb} team to have an instant view of the situation in orbit and finally get valuable clues of what happened on-board ESTCube-1 as well. The last minute changes could not help the Aalto-1 \gls{epb} to complete its mission, but at the very least they helped in compiling a road-map towards more successful missions in the future. A small step for Coulomb drag industry, but a step forward nonetheless.

\subsection{Aalto-1 Spectral Imager AaSI}

\subsubsection{In-orbit results}
\begin{figure}[h!]
    \centering
    \includegraphics[width=\columnwidth]{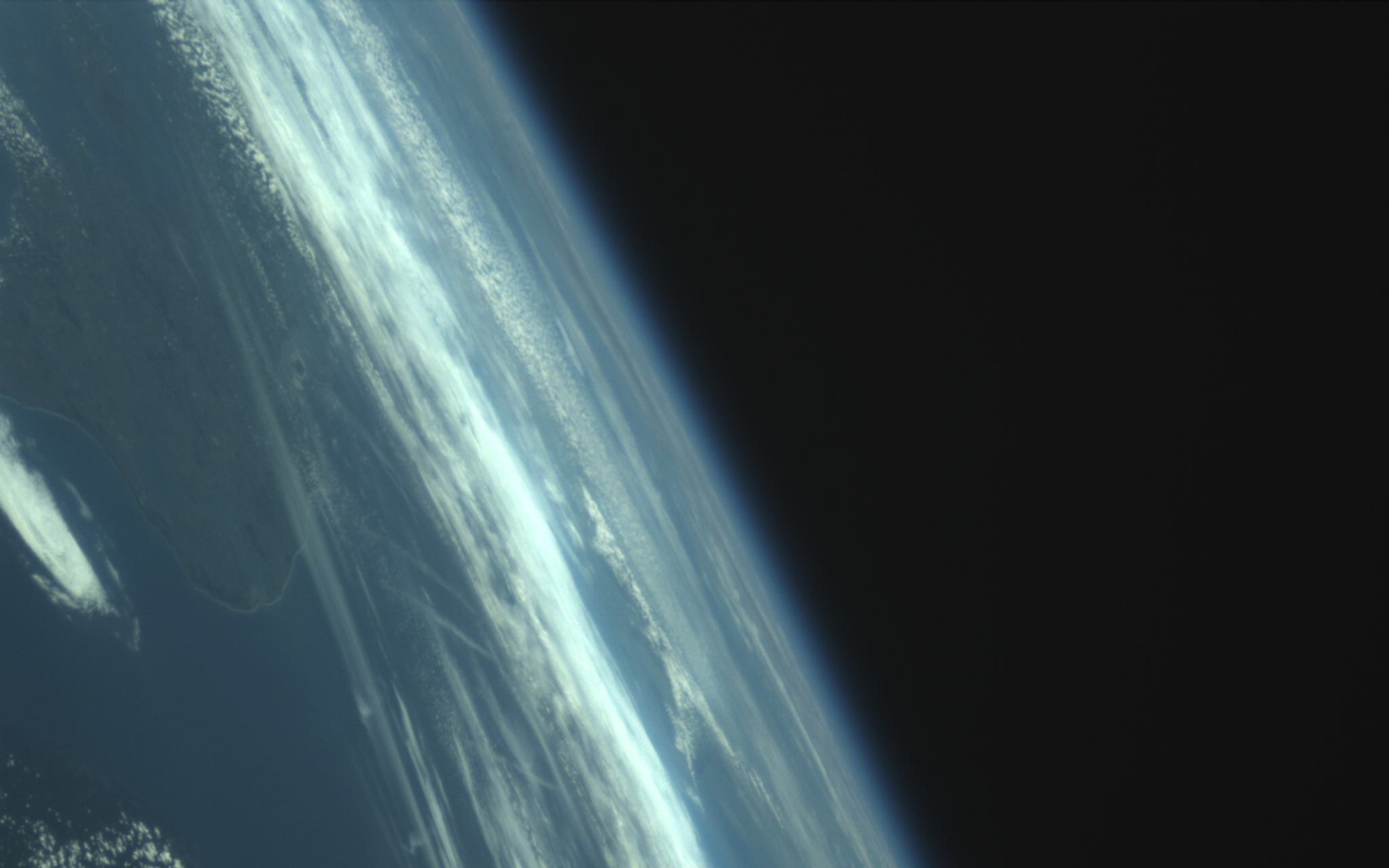}
    \caption{The first image downlinked from Aalto-1. The image is taken with the VIS camera on 5$^{th}$ of July, 2017 and it shows the coastline of Denmark together with Earth's horizon.}
    \label{fig:aasi_img1}
\end{figure}

After establishing communications, the VIS camera was first powered on the 3$^{rd}$ of July, 2017. The first housekeeping data from the camera indicated nominal behaviour, and the instrument temperature was ca.~$-5^\circ$C. The first image, as shown in Fig.~\ref{fig:aasi_img1}, was taken on the 5$^{th}$ of July, while the satellite was still tumbling. During image acquisition, the satellite was located over Norway with the field of view pointed to the southern direction towards Denmark. Based on visual analysis, the image quality is good, and no visible de-focusing or new aberrations are present.

The spectral camera was first powered on the 25$^{th}$ of July, and the instrument housekeeping data was nominal. The temperature was around $-5^\circ$C, and the piezo voltages for the FPI were between 26~V and 28~V, which indicated perfect health for the FPI unit. When compared to piezo voltages measured on ground prior to launch, there was approximately ~10~V difference in one of the channels. This was expected as there is a temperature difference between the measurements (+22$^\circ$C at the pre-launch check vs. $-5^\circ$C in orbit) and the water absorbed by the piezo actuators has evaporated at the time of taking in-orbit measurements. 

\begin{figure}[h!]
    \centering
    \includegraphics[width=0.6\columnwidth]{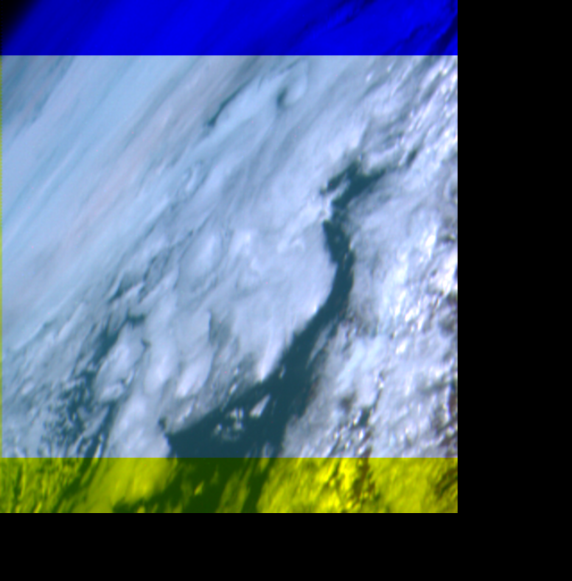}
    \caption{False color composite of the first spectral image captured by AaSI. The approximate wavelengths in the image are R=710~nm, G=535~nm and B=510~nm. The tumbling of the satellite is clearly visible as the frames do not contain much overlap. The bottom part of the image shows as yellow, as the wavelengths 535 nm and 710 nm are extracted from the same raw image.}
    \label{fig:1stSPE}
\end{figure}

First images were taken with the spectral camera on the 3$^{rd}$ of August. From these images, the functionality of the camera optics was verified. The imaged scene was covered by clouds, and in the false color composite as shown in Fig.~\ref{fig:1stSPE}, one can see spectral variation in the clouds. During this time, the satellite was still tumbling quite rapidly, so the imaged area is moving significantly between the spectral frames.

\begin{figure}[h!]
\centering
    \begin{subfigure}[b]{.5\linewidth}
    \includegraphics[width=0.7\columnwidth]{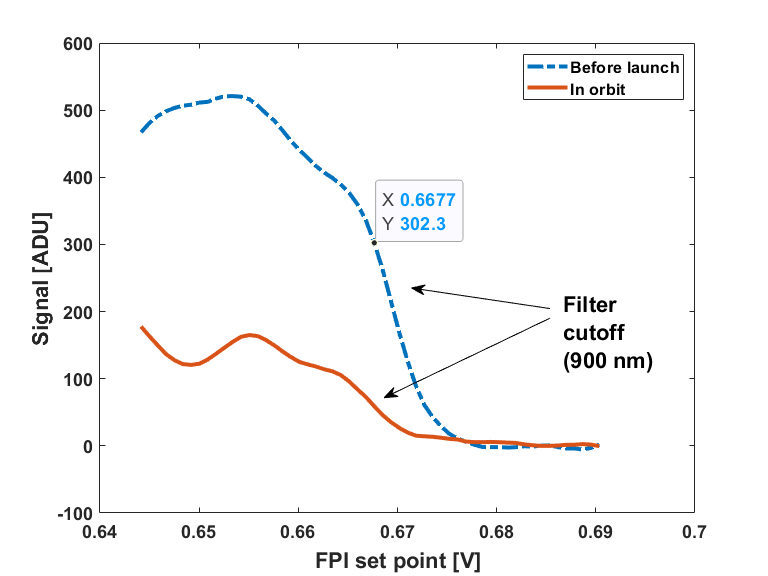} 
    \label{fig:subim1}
    \end{subfigure}
    ~
    \begin{subfigure}[b]{.5\linewidth}
    \includegraphics[width=0.7\columnwidth]{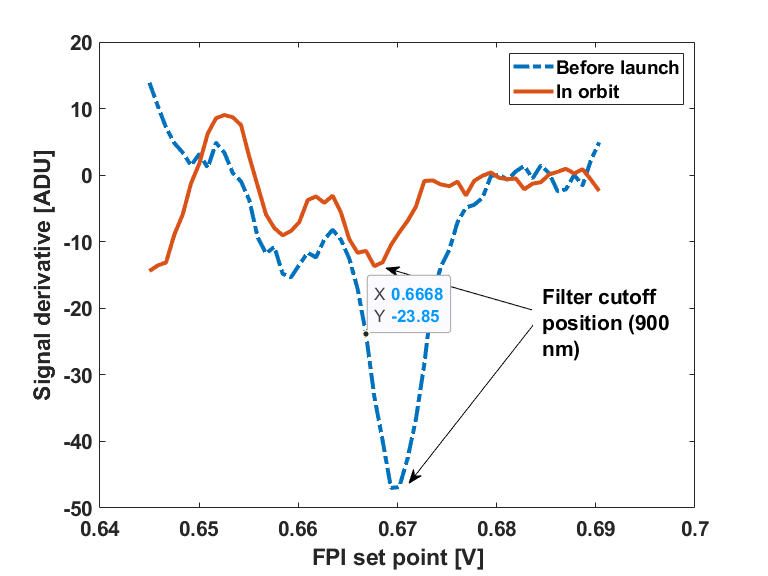}
    \label{fig:subim2}
    \end{subfigure}
\caption{Calibration measurement comparison. In the top figure, the signal level is plotted as function of FPI set point voltage. Signal derivative is plotted in the bottom figure. The position with the steepest slope corresponds to the filter cutoff wavelength. The filter cutoff position is visible in both cases, but the measurement performed in orbit is distorted. This is mainly due to the cold temperature, which is outside the instrument's operation temperature.}
\label{fig:edgecalib}
\end{figure}

After the performance of optics was verified, the on-board spectral calibration method was tested. The calibration is based on measuring a bright target (e.g. cloud or desert) and scanning the spectral filter over the cutoff wavelength of the 900 nm short pass filter and taking an average of the pixel values. The sequential images are recorded with very small wavelength increment. When the spectral transmission peak passes over the short pass filter, the signal level will drop. When the signal is plotted as a function of FPI set point voltage, the drop in signal level is visible. The location where the slope is steepest corresponds to the cutoff wavelength of the shortpass filter.  Successful calibration measurement was performed on the 5$^{th}$ of September which is shown in Fig.~\ref{fig:edgecalib}.
When compared to measurements done on ground, it can be seen that the spectral behaviour is similar, but due to the cold temperature ($-16^\circ$C) and different illumination conditions the shape of the calibration spectrum is different.

\begin{figure}[h!]
\centering 
\begin{subfigure}[b]{.5\linewidth}
\centering\includegraphics[width=0.7\columnwidth]{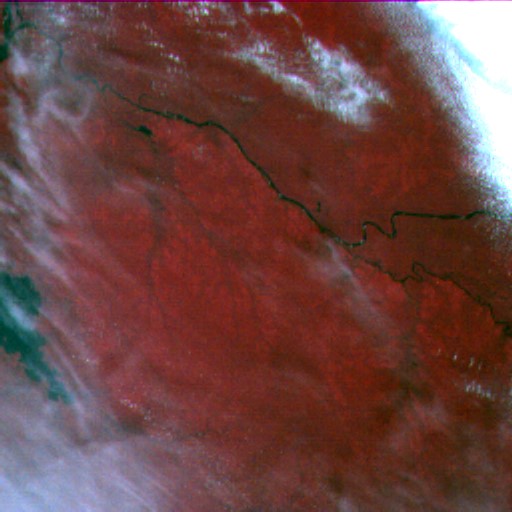}
\end{subfigure}
~
\begin{subfigure}[b]{.5\linewidth}
\centering \includegraphics[width=0.7\columnwidth]{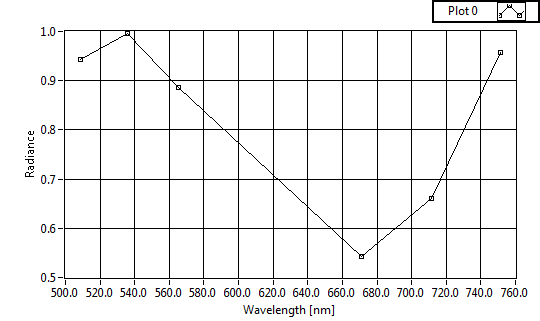}
\end{subfigure}
\caption{The first cloud-free spectral image of a land target (top). The image is centered on Tshuapa River near Mbandaka. The false color image is constructed from R=752 nm, G=671 nm, B=565 nm. The bottom figure shows the spectrum of the central area of the image measured at 6 wavelenghts.} 
\label{fig:cloud_free_spe}
\end{figure}

The satellite was de-tumbled in June 2018 and the imaging campaign was continued immediately after de-tumbling. During this campaign, an image mosaic was created from VIS images, and finally on  August 6, 2018 the first cloud-free images of land targets were acquired. The imaging sequence started at the equator above Congo, and continued for about six minutes while the satellite was travelling south toward South Africa.
The images of six different wavelengths were acquired, and, from the resulting spectrum, the red-edge of vegetation is clearly visible, as shown in Fig.~\ref{fig:cloud_free_spe}.

\begin{figure}[h!]
    \centering
    \includegraphics[width=0.9\columnwidth]{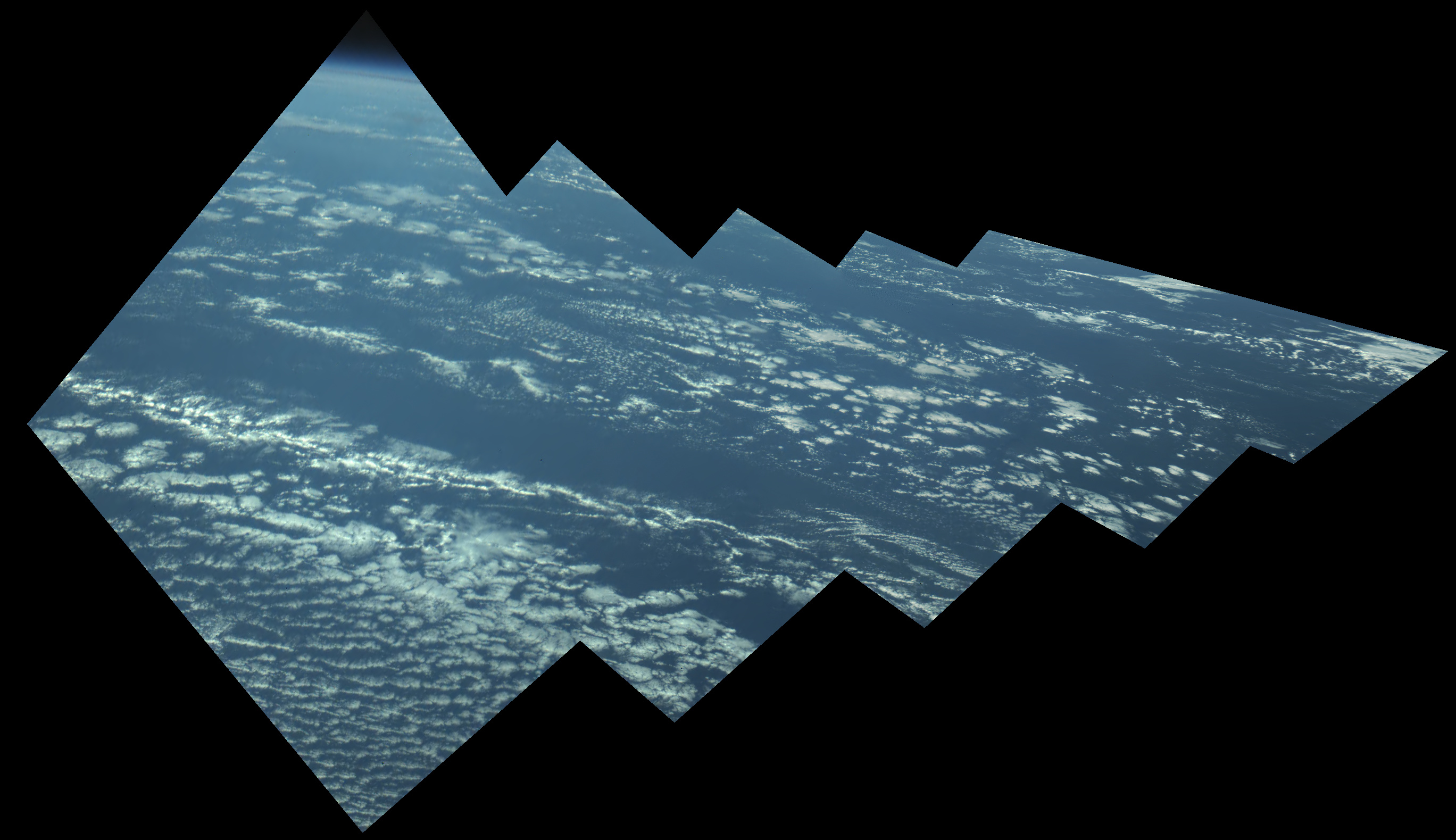}
    \caption{Mosaic of sequential VIS images}
    \label{fig:mosaic}
\end{figure}

An image compression program was uploaded to the satellite during the spring of 2018. This was first tested around the midsummer of 2018, and several series of images were taken. In order to downlink the image mosaics, image compression was required. After compression, the images were successfully downlinked. The stiched mosaic is shown in Fig.~\ref{fig:mosaic}. The slow tumbling of the satellite is clearly visible in the sequential images.

\subsubsection{Lessons learned}
This was the first mission ever to demonstrate a hyperspectral camera on a nanosatellite. It was also the first space-borne demonstration of a tunable \gls{fpi}-based nanosatellite-compatible hyperspectral camera. 

The main lesson learned was that this technology works in space environment and it can be used for nanosatellite-based hyperspectral imagers. All of the primary mission objectives were completed, so the AaSI mission can be considered successful.

Not all functionalities of the imager could be verified though. The tumbling platform prevented imaging of planned targets, and the limited downlink allowed only the use of minimal spectral mode with six wavelengths. However, the tumbling platform showcased the benefits of frame-based spectral imaging, as the images in different wavelengths can be overlapped in post processing. This is a great benefit in nanosatellite missions, as the imager can still be used in the case of attitude control malfunction.

\section{Platform in-orbit performance}
\label{sec:platform}

The in-orbit performance of spacecraft platform which consisted of commercial and in house developed subsystems, is briefly presented. While the key platform subsystems were successfully commissioned, the spacecraft accomplished its mission with partial success. The design approach of platform subsystems which consisted of an \gls{eps} \cite{Hemmo13Master}, an \gls{adcs} \cite{Tikka14Attitude}, a \gls{gps}-based navigation system \cite{Leppinen13Master}, a \gls{uhf} \cite{Lankinen15Master} and S-band \cite{Jussila13SBand} radios for \gls{tt&c}, and a Linux-based \gls{obdh} \cite{Leppinen2019OBC} is briefly presented in \cite{a1_technology_2020}.

\subsection{In orbit performance of \gls{eps}}

\begin{figure}[h!]
    \centering
    \includegraphics[width=\columnwidth]{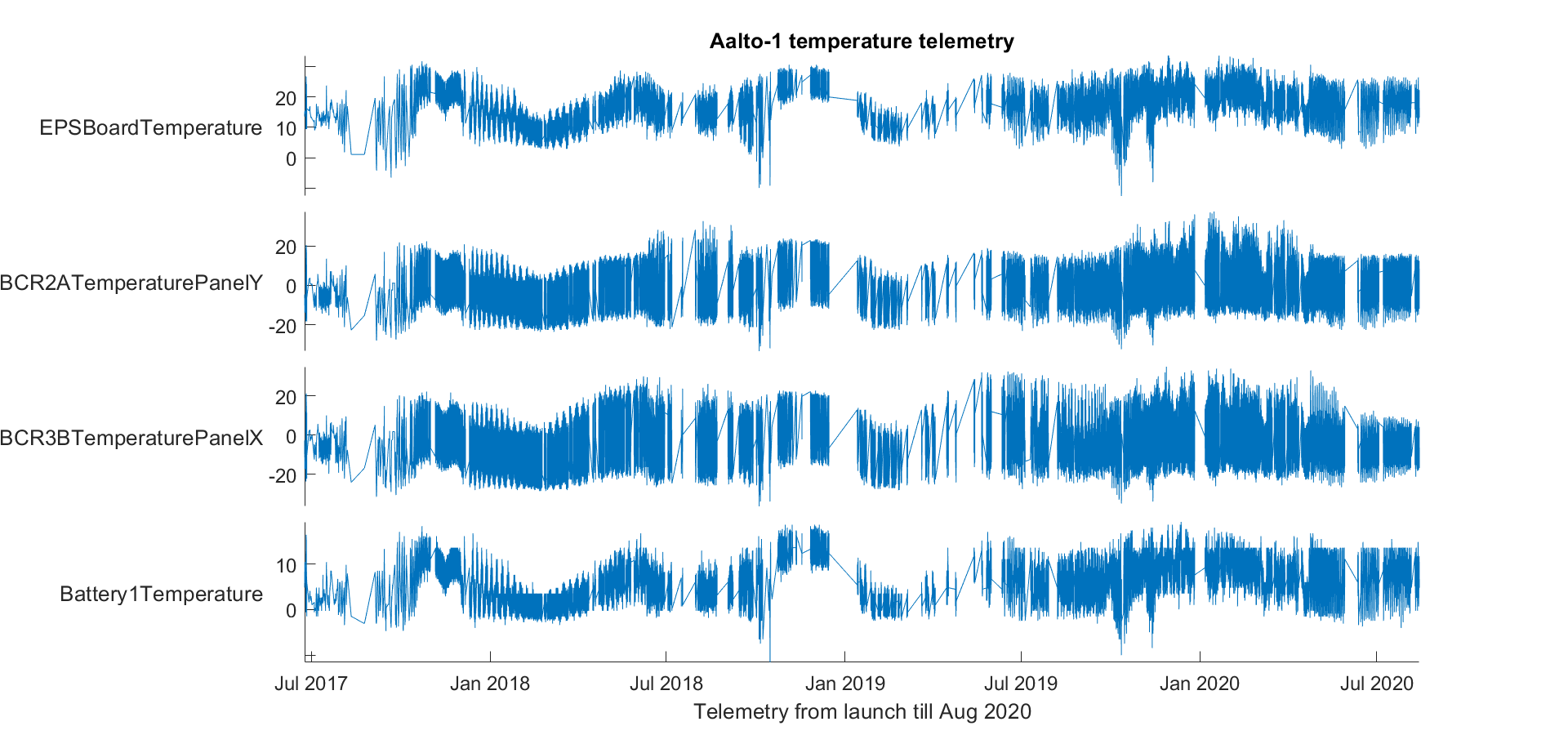}
    \caption{Aalto 1 surface and inner temperatures (in ${\circ}$C ) from launch till Aug 2020}
    \label{fig:a1_temp}
\end{figure}

\begin{figure}
    \centering
    \includegraphics[width=\columnwidth]{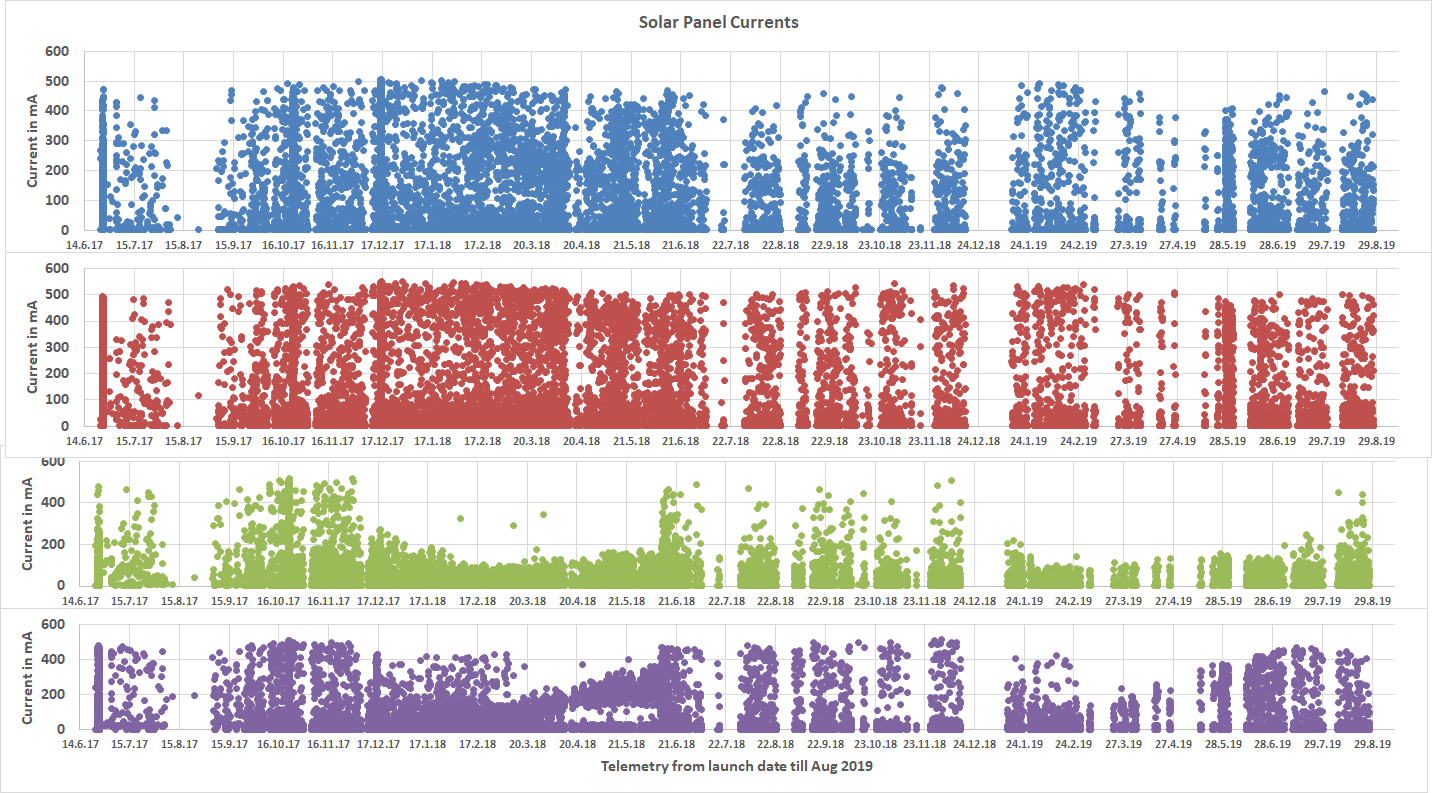}
    \caption{Solar panel current intensities from launch till Aug 2019. The vertical axis represents the generated current (in mA) read by each BCR}
    \label{fig:eps1}
\end{figure}

In order to monitor and keep track of the health of the spacecraft, a number of housekeeping sensors were used. The performance of \gls{eps} is presented in terms of telemetry values of voltage, current and temperature sensors. The flight data of these sensors confirms that the \gls{eps} is functional and provides power to satellite subsystems since its launch. However, there are some issues. The telemetry data reveals partially degraded performance of one of the solar panels as evident by green plot in Fig.~\ref{fig:eps1}. This behaviour is likely due to the un-controlled spin orientation of the satellite.
The telemetry data of solar panel temperatures, \gls{eps} board temperature and battery temperatures  from launch date till Aug 2020 is plotted in Fig.~\ref{fig:a1_temp}.
The highest temperature variation takes place on the satellite surface as evident from central graphs representing panel X and Y, where solar panel temperatures change in $\pm$20$^{\circ}$C range. This range remains quite stable throughout the mission representing that the temperature is at equilibrium. The temperature inside the satellite depends on operation of payloads and platform subsystems. The telemetry data of board and battery temperatures, as evident from Fig.~\ref{fig:a1_temp}, represents that the passive thermal control maintains sufficiently stable  temperature fluctuations.

\subsection{In-orbit performance of \gls{adcs}}

\begin{figure}[h!]
    \centering
    \includegraphics[width=\columnwidth]{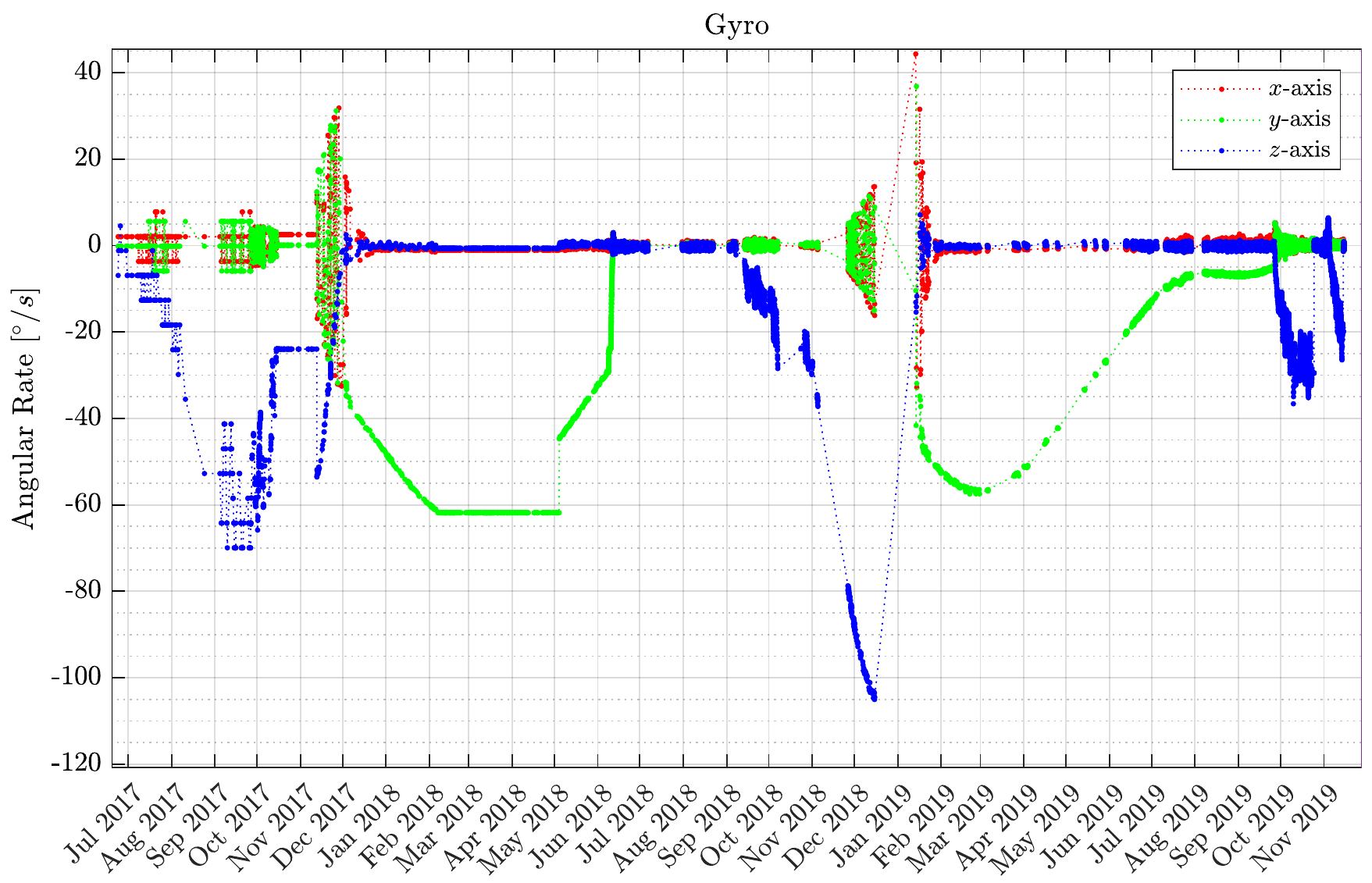}
    \caption{Gyroscope data from July 2017 to November 2019.}
    \label{fig:adcs1}
\end{figure}

The commissioning phase of the \gls{adcs} functions were met with complications, as some of the sensor readings were erroneous. Two of the sun sensors (on the +X and $-$X directions of the satellite) were malfunctioned and not usable for attitude estimation. In Fig.~\ref{fig:adcs1}, the gyroscope readings from regular housekeeping data until October 2017 have a low angular rate resolution. This was because of improper processing of sensor raw data and a problem with the communication channel in the \gls{adcs} module. This was fixed with a small firmware update. Over the course of the mission, sometimes the \gls{adcs} module is reset and defaulted to idle mode. Such event will turn off the ADCS sensors which needs to be manually turned on. This shows up as frozen sensor data, visible in Fig.~\ref{fig:adcs1} around October--November, 2017 and February--May 2018.

The first attempt to detumble the satellite was attempted on October 2017. The attempt failed because of constant rebooting of the \gls{adcs} module when the B-dot control was turned on. The problem was caused by the magnetorquer driver channel which  was later fixed with a major firmware and magnetorquer driver update on June 2018 consequently solving the reboot issue. The detumbling operation was tested again with positive results. The spin rate of the satellite was reduced close to 0~deg/sec as confirmed by the telemetry data of Fig.~\ref{fig:adcs1}. The detumbling control was kept on until September 2018, after which the satellite started to spin up.

The main cause of the uncontrolled spin up of the satellite remains unknown when the B-dot is disabled. Some possible causes are environmental disturbance torque, residual dipole moment generated by unknown magnetic materials or current loop from the solar panels power routing.

Although detumbling with the B-dot controller was successful, many other mission modes, including controlled spin up for tether deployment, were not successful. The \gls{adcs} commissioning modes, including the spin up manoeuvre, has been tried with the magnetorquers only \cite{mughal, mukhtar}. The reaction wheels showed inconsistencies in their power reading during the early commissioning phase and thus have not been thoroughly tested yet.

An important lesson learned was to procure the commercial modules at the early stages of development and  test all the functional modes during the qualification phase. Moreover, designing an in house subsystem gives more flexibility in interfacing and testing.

\subsection{In-orbit performance of \gls{obdh}, \gls{tt&c} and \gls{gps} subsystems}

\begin{figure}[h!]
    \centering
    \includegraphics[width=\columnwidth]{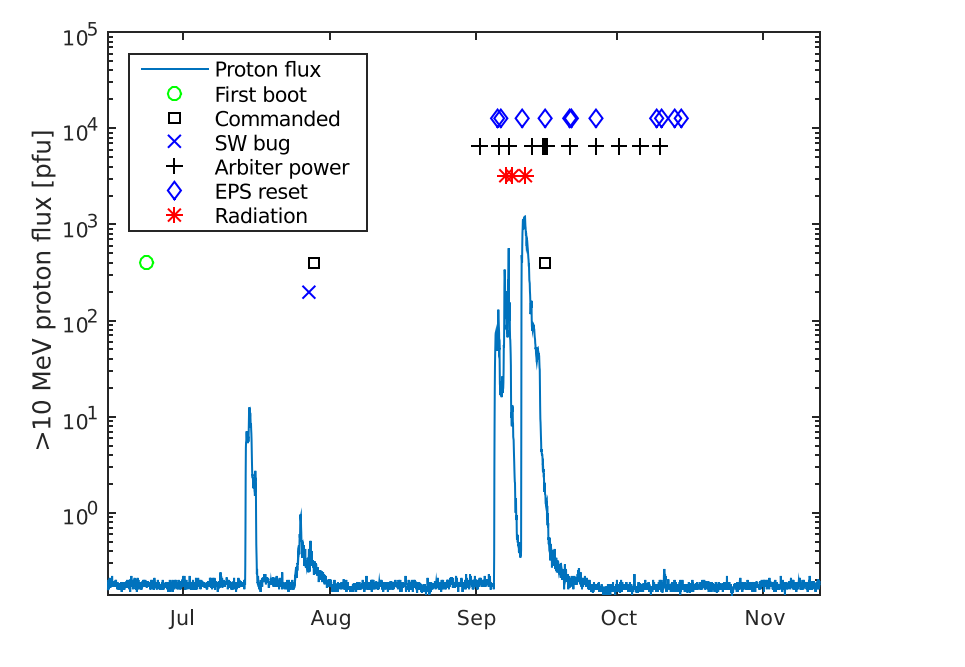}
    \caption{\gls{obc} reboot events during July-Nov 2017 \cite{Leppinen2019OBC}}
   \label{fig:obc_reboots}
\end{figure}

The \gls{obc}$-$1 branch that was enabled at 
satellite deployment had reset itself after around a month after launch. The cause was initially perceived  as single event upset due to radiation in South Atlantic Anomaly, but the problem was later found to be actually caused by a software bug in the command line. The reboots due to software bugs, radiation and \gls{eps} reset etc. plotted in relation to proton fluxes, are given in Fig.~\ref{fig:obc_reboots} \cite{Leppinen2019OBC}.
During first five months after launch, a total of 38 boot events occurred.  A boot event may have one or more boots, with the group having a likely common cause. A further detailed analysis on the boot events can be read in \cite{Leppinen2019OBC}.
The satellite suffered from instability in \gls{eps}, resulting in several resets in \gls{eps} and the arbiter. A Coronal Mass Ejection (CME) occurred in early September 2017, providing an excellent opportunity for \gls{radmon} testing \cite{Vainio2018PhysicsDays}. The satellite was quickly retasked to collect as much data as possible with \gls{radmon}. A precious \gls{radmon} set of data collection was also interrupted during a CME due to \gls{obc} reboots. A few unexplained boot events of the \gls{obc}, which were resolved without involvement of the arbiter, occurred during the CME event. It is suspected that these may be related to either radiation or \gls{eps} reset.

Immediately after the launch, multiple objects launched on the same rocket as Aalto-1 had similar \gls{tle}, and it was unclear which \gls{tle} set belonged to Aalto-1. The GPS subsystem was one of the first instruments successfully operated after contacting the satellite, and navigation solutions provided by the receiver allowed determining the correct \gls{tle} set. The determined identity was also communicated to the \gls{tle} data provider \cite{Leppinen18PhD}.

It has been observed that the \gls{tle} accuracy has been sufficient for most routine operations, and the use of GPS has been less frequent than expected. A sub-optimal GPS antenna placement (resulting from a compromise with solar panel placement) and satellite tumbling have caused delays in obtaining the first fix after powering the receiver.

The commissioning of the \gls{uhf} transceiver was successful since the first contact with the CubeSat was established during first pass over the ground station. The commissioning phase was met with many challenges which have been briefly detailed in \cite{a1_technology_2020}. From the telemetry logs, a radio interference in the Northern direction, close to the horizon, was noted at around 437.22~MHz. Similar kind of interference around the 437.0--437.4~MHz was measured by the UWE-3 CubeSat mission though the source of interference has not been confirmed \cite{UWE3_interference}. The S-band transmitter has not been successfully commissioned despite multiple attempts in July 2017 and July 2018. 

\newpage

\section{Discussion and conclusions}
\label{sec:discussion}

Although the mission was a partial success in terms of executing the experiments, the important lessons learned during this mission have been applied in the design of next variants of payloads and platforms. The \gls{radmon} instrument was successful in commissioning and measurement phases. Its heritage has been used to design a more complex \gls{pate} payload for the upcoming FORESAIL-1 mission~\cite{Oleynik2020ParticlePerformance}. The \gls{epb} tether could not be deployed due to a failure in tether deployment hardware. The lessons learned have been taken into consideration in development of the plasma brake for upcoming FORESAIL-1 and ESTCube-2 missions~\cite{Iakubivskyi2020}. The \gls{aasi} was the first nanosatellite-compatible hyper-spectral imager to be flown in space. Aalto-1 project successfully demonstrated the expertise of VTT in both visible and hyper-spectral miniature imager designs. The technology has many  potential future applications to serve CubeSat and/or scientific industry/community. Since Aalto-1, VTT's hyper-spectral imagers have been developed for Reaktor Hello World, PICASSO, Hera and Comet Interceptor missions. The platform has provided successful in-orbit demonstration, although some subsystems lacked the desired performance. An important lesson learned was to perform a rigorous test campaign while integrating the commercial and in-house built subsystems.

\section*{Acknowledgements} 
The RADMON team thanks P.-O.\ Eriksson and S.\ Johansson at the Accelerator Laboratory, \AA{}bo Akademi University, for operating the cyclotron.
Computations necessary for the presented modeling were conducted on the Pleione cluster at the University of Turku.

Aalto University and its Multidisciplinary Institute of Digitalisation and Energy are thanked for Aalto-1 project funding, as are Aalto University, Nokia, SSF, the University of Turku and RUAG Space for supporting the launch of Aalto-1.

\bibliographystyle{model1-num-names}
\bibliography{ms}
\end{document}